\documentstyle[12pt]{article}
\makeatletter
\def\slash#1{{\mathpalette\c@ncel{#1}}} 
\makeatother

\newcommand{\ub}{\bar u}
\newcommand\beq{\begin{eqnarray}}
\newcommand\eeq{\end{eqnarray}}
\newcommand\be{\begin{eqnarray}}
\newcommand\ee{\end{eqnarray}}
\newcommand\la{\langle}
\newcommand\ra{\rangle}
\def\Mom{{\cal M}}

\def\kslash{\slash{\mkern-1mu k}}

\def\ieps{i\varepsilon}
\def\pp{{\bf p}_\perp}
\def\kp{{\bf k}_\perp}

\begin{document}


\begin{titlepage}
\begin{flushright}
\begin{tabular}{l}
\end{tabular}
\end{flushright}
\vskip0.5cm
\begin{center}
  {\Large \bf
             Violation of Sum Rules for Twist-3 Parton Distributions
             in QCD
  \\}

\vspace{1cm}
{\sc Matthias~Burkardt}${}^1$ and
 {\sc Yuji~Koike}${}^2$
\\[0.3cm]
\vspace*{0.1cm} ${}^1${\it Dept. of Physics, New Mexico State University,
Las Cruces, NM 88003, USA}
\\[0.3cm]
\vspace*{0.1cm} ${}^2$ {\it Dept. of Physics, Niigata University,
Niigata 950--2181, Japan}
\\[1cm]

{\em Version of \today}\\[1cm]

  \vskip1.8cm
  {\large\bf Abstract:\\[10pt]} \parbox[t]{\textwidth}{ 
Sum rules for twist-3 distributions are reexamined.
Integral relations between twist-3 and twist-2
parton distributions suggest the possibility for
a $\delta$-function at $x=0$. We confirm and clarify
this result by constructing $h_L$ and $h_L^3$ (quark-gluon interaction
dependent part of $h_L$) 
explicitly from their moments for a one-loop
dressed massive quark.
The physics of these results is illustrated
by calculating $h_L(x,Q^2)$ using light-front 
time-ordered pQCD to ${\cal O}(\alpha_S)$ on a quark target. 
A $\delta(x)$ term is also found in $e(x,Q^2)$,
but not in $g_T(x,Q^2)$, to this order in ${\cal O}(\alpha_S)$.
}
  \vskip1cm 
\end{center}
\end{titlepage}

\section{Introduction}
\setcounter{equation}{0}

Ongoing experiments with polarized beams and/or targets
conducted at RHIC, HERMES and COMPASS etc
are providing us with important information on the spin distribution
carried by quarks and gluons in the nucleon.
They are also enabling us to extract information on the
higher twist distributions which represent the effect of
quark-gluon correlations.  
In particular, the twist-3 distributions $g_T(x,Q^2)$ and $h_L(x,Q^2)$
are unique in that they appear as a leading contribution
in some spin asymmetries:  For example, $g_T$ can be measured in the 
transversely polarized lepton-nucleon deep inelastic scattering
and $h_L$ appears in the longitudinal-transverse spin asymmetry
in the polarized nucleon-nucleon Drell-Yan process\,\cite{JJ92}.
The purpose of this paper is to reexamine the
validity of the sum rules for these twist-3 distributions.

The complete set of the twist-3 quark distributions in our interest
are given as the light-cone corelation functions in a
parent hadron with momentum $P$, spin $S$ and mass $M$:
\begin{eqnarray}
& &\int {d\lambda\over 2\pi}e^{i\lambda x}
\langle PS|\bar{\psi}(0)\gamma^\mu\gamma_5\psi(\lambda n)
|_{Q^2}|PS\rangle
\nonumber\\
& &\qquad=
2\left[g_1(x,Q^2)p^\mu (S\cdot n) 
+ g_T(x,Q^2)S_\perp^\mu +M^2 g_3(x,Q^2)n^\mu (S\cdot n)
\right],
\label{eq1}\\
& &\int{d\lambda\over 2\pi}e^{i\lambda x}
\langle PS | \bar{\psi}(0)\sigma^{\mu\nu}i\gamma_5 \psi(\lambda n)|_{Q^2}
|PS \rangle
=2\left[h_1(x,Q^2)(S_\perp^\mu p^\nu - S_\perp^\nu p^\mu )
/M\right.\nonumber\\
& &\left.\qquad\qquad
+h_L(x,Q^2)M(p^\mu n^\nu - p^\nu n^\mu )(S\cdot n)
+h_3(x,Q^2) M (S_\perp^\mu n^\nu - S_\perp^\nu n^\mu)\right],
\label{eq2}\\
& &\int{d\lambda\over 2\pi}e^{i\lambda x}
\langle PS|\bar{\psi}(0)\psi(\lambda n)|_{Q^2}|PS\rangle
=2Me(x,Q^2)\label{eq3},
\end{eqnarray}
where two light-like vectors $p$ and $n$
are introduced by the relation
$p^2=n^2=0$, $n^+=p^-=0$, $P^\mu = p^\mu +{M^2\over 2}n^\mu$ and
$S^\mu$ is decomposed as
$S^\mu=(S\cdot n)p^\mu + (S\cdot p)n^\mu + S_\perp^\mu$.
The variable $x$ represents the parton's light-cone 
momentum fraction and each function has a support 
for $x$ on $[-1,1]$.  The anti-quark distributions 
$\bar{g}_{1,T,3}(x,Q^2)$, $\bar{h}_{1,L,3}(x,Q^2)$ 
are obtained by the replacement of $\psi$ into its 
charge
conjugation field $C\bar{\psi}^T$ in (\ref{eq1})-(\ref{eq3})
and are related to the quark distributions as
$\bar{g}_{1,T,3}(x,Q^2) = {g}_{1,T,3}(-x,Q^2)$ and
$\bar{h}_{1,T,3}(x,Q^2) = - {h}_{1,T,3}(-x,Q^2)$.  
The sum rules in our interest are obtained by taking the first
moment of the above relations.  For example, from (\ref{eq1}), one obtains
\begin{eqnarray}
& &\langle PS |\bar{\psi}(0)\gamma^\mu\gamma_5\psi(0)|_{Q^2}|PS\rangle
\nonumber\\
& &\qquad\qquad=2\int_{-1}^1\,dx\,\left[ g_1(x,Q^2)p^\mu(S\cdot n)+
g_T(x,Q^2)S_\perp^\mu + M^2 g_3(x,Q^2)n^\mu(S\cdot n)\right].
\label{eq4}
\end{eqnarray}
{}From rotational invariance, it follows that
the left hand side of (\ref{eq4}) is proportional to the spin vector
$S^\mu$ and thus $g_{1,T,3}(x,Q^2)$ must satisfy
\begin{eqnarray}
\int_{-1}^1\,dx\,g_1(x,Q^2) = \int_{-1}^1\,dx\,g_T(x,Q^2),
\label{eq5}\\
\int_{-1}^1\,dx\,g_1(x,Q^2) = 2\int_{-1}^1\,dx\,g_3(x,Q^2).
\label{eq6}
\end{eqnarray}
The same argument for (\ref{eq2}) leads to the sum rule relations
for $h_{1,L,3}(x,Q^2)$:
\begin{eqnarray}
\int_{-1}^1\,dx\,h_1(x,Q^2) = \int_{-1}^1\,dx\,h_L(x,Q^2),
\label{eq7}\\
\int_{-1}^1\,dx\,h_1(x,Q^2) = 2\int_{-1}^1\,dx\,h_3(x,Q^2).
\label{eq8}
\end{eqnarray}
The sum rule (\ref{eq5}) is known as 
Burkhardt-Cottingham sum rule\,\cite{BC}
and (\ref{eq7}) was first
derived in Refs. \cite{mb:nagoya,TM94,Bur95}.  
Since the twist-4 distributions $g_3$, $h_3$ are unlikely to be measured
experimentally,
the sum rules involving those functions
(\ref{eq6}) and (\ref{eq8}) are practically
useless and will not be addressed in the subsequent discussions.  
Since the left hand side of (\ref{eq5}) and (\ref{eq7}) are, respectively,
the axial charge and the tensor charge, the integral
itself in the right hand side of (\ref{eq5})-(\ref{eq8}) are finite.  
As is clear from the above derivation,
these sum rules are mere consequences of the rotational invariance
and there is no doubt in its validity in mathematical sense.
However, if one try to confirm those sum rules by experiment,
a great care is required to perform the integral including $x=0$.  
For example, in DIS, $x$ is identified as the Bjorken's
variable $x_B=Q^2/2P\cdot q$ and $x=0$ corresponds
to $P\cdot q\to \infty$ and this limit can never be achieved
in the rigorous sense.  
Accordingly, if $h_L(x,Q^2)$ has 
a contribution proportional to $\delta(x)$ and 
$h_1(x,Q^2)$ does not,
experimental measurement can never confirm the
sum rule (\ref{eq7}) but would rather claim the violation of the sum rule.
Such a contribution has already been suggested in 
Refs. \cite{mb:nagoya,Bur95}. 
In Refs. \cite{JMR97,DB98,Bur99} it
was also argued that the integral of 
$h_L(x,Q^2)-h_1(x,Q^2)$ 
can be related to the value of
certain `surface terms' that appear in formal
manipulations involving integrations by parts.

The purpose of this paper is to reexamine the 
sum rules involving the first moment of the twist-3
distributions.  In particular we will argue that
the twist-3 distribution $h_L(x,Q^2)$ has a potential $\delta$-singularity
at $x=0$, assuming that the twist-2 distributions
$g_1(x,Q^2)$ and $h_1(x,Q^2)$ do not have such singularity.
The paper is organized as follows:  
In section 2, we examine the sum rule for $h_L$.
Starting from the general decomposition of $h_L$ based on the QCD
equation of motion, we will show that it contains a
function $h_L^m$ which has $\delta(x)$-singularity (Section 2.1).  
In Section 2.2, we construct $h_L$ for a massive quark
>from the moment of $h_L^3$ in the one-loop level
and show that the $h_L^3$ also has an $\delta(x)$-singularity,
which together with the singularity in $h_L^m$ will give 
rise to a $\delta(x)$ singularity in $h_L$ itself. 
In section 2.3, we will perform an explicit light-cone 
calculation of $h_L$ in the one-loop level to confirm the 
result in the previous sections.
Sections 3 and 4 are, respectively, devoted to similar
examination for the sum rules for $e(x,Q^2)$ and $g_T(x,Q^2)$.

\section{$h_L(x,Q^2)$ for a massive quark}
\setcounter{equation}{0}

\subsection{$\delta(x)$-functions in $h_L(x,Q^2)$}

The OPE analysis of the correlation function (\ref{eq2}) allows us to
decompose $h_L(x,Q^2)$ into the contribution 
expressed in terms of twist-2 distributions
and the rest which we call $h_L^3(x,Q^2)$.
Since the scale dependence of each distribution
is inessential in the following discussion, we shall omit it in this 
subsection
for simplicity.  
Introducing the notation for the moments on $[-1,1]$,
\beq
{\cal M}_n[h_L]\equiv \int_{-1}^1\,dx\,x^nh_L(x),
\eeq
this decomposition is given in terms of the moment relation\,\cite{JJ92}:
\beq
\Mom_n[h_L] &=& {2\over n+2}\Mom_n[h_1] + {n\over n+2}{m_q\over M}\Mom_{n-1}[g_1]
+\Mom_n[h_L^3],\qquad (n\geq 1)
\label{eq:hlmomn}\\
\Mom_0[h_L] &=& \Mom_0[h_1],
\label{eq:hlmom0}
\eeq
with the condition 
\beq
\Mom_0[h_L^3]=0,
\label{eq:hl3mom0}\\
\Mom_1[h_L^3]=0.
\label{eq:hl3mom1}
\eeq
By inverting the moment relation, one finds
\beq
h_L(x) &=& h_L^{WW}(x) + h_L^m(x) + h_L^3(x)
\label{eq:ope1a}\\
&=& \left\{
\begin{array}{l}
\displaystyle{
2x\int_x^1dy \frac{h_1 (y)}{y^2}
+\frac{m_q}{M} \left[ \frac{g_1(x)}{x} -2x \int_x^1 dy
  \frac{g_1(y)}{y^3}\right] + h_L^3(x) }\\[10pt]
\qquad\qquad\qquad\qquad\qquad\qquad\qquad\qquad\qquad (x>0) \\[12pt] 
\displaystyle{
 -2x\int_{-1}^xdy \frac{h_1 (y)}{y^2}
+\frac{m_q}{M} \left[ \frac{g_1(x)}{x} +2x \int_{-1}^x dy
  \frac{g_1(y)}{y^3}\right] + h_L^3(x),
} \\[10pt]
\qquad\qquad\qquad\qquad\qquad\qquad\qquad\qquad\qquad (x<0) 
\end{array}
\right.
\label{eq:ope1b}
\eeq
where the first and second terms in Eq. 
(\ref{eq:ope1a}) denote the corresponding terms 
in (\ref{eq:ope1b}).
In this notation the sum rule (\ref{eq:hlmom0}) and the condition
(\ref{eq:hl3mom0}) implies\,
\footnote{More precisely, the original OPE tells us
${\cal M}_0[h_L^3+h_L^m]=0$.   But as long as $g_1(0_\pm)$ is finite,
which we will assume, this is equivalent to
stronger relations (\ref{eq:hl3mom0}) and (\ref{eq:hlm0}).}
\beq
{\cal M}_0[h_L^m]=0.
\label{eq:hlm0}
\eeq
If one naively
integrates 
(\ref{eq:ope1b}) 
over $x$ at $x>0$ and $x<0$, 
while dropping all surface terms one
arrives at $\int_0^1dx h_L(x)=\int_0^1 dx h_1(x)
+\int_0^1 dx h_L^3(x)$ and likewise for $\int_{-1}^0 dxh_L(x)$.
This result, in combination with (\ref{eq:hl3mom0}),
implies that
$\int_{-1}^1 dx h_L=\int_{-1}^1 dx h_1$.
In the following we will argue that this procedure
may be wrong due to the potentially very
singular behavior of the functions involved
near $x=0$. Investigating this issue in detail  
will be the main purpose of this paper.

We first address the potential singularity at $x=0$ in the integral
expression for $h_L^m(x)$ in (\ref{eq:ope1b}).    
In order to regulate the region near $x=0$, we first
multiply $h_L^m(x)$
by $x^\beta$, integrate from $0$ to $1$
and let $\beta \rightarrow 0$. This yields
\beq
\int_{0+}^1 dx h_L^m(x) 
= \frac{m_q}{2M}\lim_{\beta \rightarrow 0} \beta \int_0^1 dx x^{\beta -1}
g_1(y)
={m_q\over 2M} g_1(0+),\nonumber\\ 
\label{eq:ope2a}
\eeq 
while multiplying Eq. (\ref{eq:ope1b}) by $|x|^\beta$ and integration from
$-1$ to $0$ yields
\beq
\int_{-1}^{0-} dx h_L^m(x)
= -\frac{m_q}{2M}\lim_{\beta \rightarrow 0} \beta \int_{-1}^{0-} dx 
|x|^{\beta -1}
g_1(y)
=-{m_q\over 2M} g_1(0-),\nonumber\\ 
\label{eq:ope2b}
\eeq 
where we have assumed that $g_1(0+)$ and $g_1(0-)$ are finite.  
Adding these results we have
\beq
\int_{-1}^{0-}dx h_L^m(x)+ \int_{0+}^1dx h_L^m(x) =
{m_q\over 2M} \left(g_1(0+)-g_1(0-)\right).
\label{eq:hlmsing}
\eeq
Since we have the sum rule (\ref{eq:hlm0})
and, in general, $\lim_{x\rightarrow 0}
g_1(x) - g_1(-x)\neq 0$, \footnote{For example,
dressing a quark perturbatively at ${\cal O}(\alpha_S)$ yields
$g_1(0+)\neq 0$ and $g_1(0-)\equiv \bar{g}_1(0+)=0$.}
we are lead to conclude
\beq
h_L^m(x)=h_L^m(x)_{reg} -{m_q\over 2M} \left(g_1(0+)-g_1(0-)\right)
\delta(x),
\label{eq:hlmreg}
\eeq
where $h_L^m(x)_{reg}$ stands for the part which is defined 
by the integral in (\ref{eq:ope1b}) at $x>0$ and $x<0$ and is regular
at $x=0$.  
Since it is unlikely that $h_L^{WW}(x)$
contains a $\delta$-function at the origin, 
the relation (\ref{eq:hlmreg}) indicates
that $h_L$ has a $\delta(x)$ term unless $h_L^3(x)$ has a 
$\delta(x)$ term and it cancels the above 
singularity in $h_L^m(x)$.

Equation (\ref{eq:hlmreg}) 
clearly demonstrate that 
the functions constituting $h_L(x)$ 
are more singular
near $x=0$ than previously assumed and that great
care needs to be taken when replacing integrals
over nonzero values of $x$ by integrals that 
involve the origin. 
In particular,
if it turns out that $h_L(x)$
itself contains a $\delta(x)$ term, then 
(\ref{eq:hlmom0}) implies
\beq
\int_{0+}^1\left(h_L(x)-h_1(x)\right)
+\int_{-1}^{0-}\left(h_L(x)-h_1(x)\right)
\neq 0,
\eeq
since $h_1(x)$ is free from singularity at $x=0$:
\beq
\int_{-1}^{0-}dx h_1(x)+ \int_{0+}^1dx h_1(x) = 
\int_{-1}^1 dx h_1(x).
\label{eq:h10}
\eeq
Accordingly an attempt to verify the ``$h_L$-sum 
rule'' \cite{mb:nagoya}
would obviously fail.

However, as we mentioned earlier, 
in order to see whether the 
$\delta(x)$ identified in (\ref{eq:hlmreg}) 
eventually survives or not in 
$h_L(x)$, we have to investigate the behavior of $h_L^3(x)$ at $x=0$.  
To this end 
we will explicitly construct $h_L(x)$ for a massive
quark to ${\cal O}(\alpha_S)$.

\subsection{$h_L(x,Q^2)$ from the moment relations}
In this subsection we will construct 
$h_L(x,Q^2)$ for a massive quark to ${\cal O}(\alpha_S)$
{}from the one-loop calculation of ${\cal M}_n[h_L^3]$.  

One-loop calculation for a massive quark (mass $m_q$)
gives $h_L(x,Q^2)$ in the following form:  
\beq
h_L(x,Q^2)=h_L^{(0)}(x) + {\alpha_S\over 2\pi}C_F{\rm ln}{Q^2\over m_q^2}
h_L^{(1)}(x),
\label{eq:hl(1)}
\eeq
where the scale $Q^2$ is introduced as an ultraviolet cutoff and
the $C_F=4/3$ is the color factor. 
$h_L^{WW,3,m(0,1)}(x)$ are defined similarly.  
$g_1^{(0)}(x)=h_1^{(0)}(x)=\delta (1-x)$ gives 
$h_L^{(0)}(x)=\delta(1-x)$, as it should.
One loop calculation for $g_1(x)$ and $h_1(x)$ 
for a quark yields
the well known splitting functions\,\cite{DGLAP,AM90}:
\beq
g_1^{(1)}(x)&=&{1+x^2 \over [1-x]_+} + 
{3\over 2}\delta(1-x),
\label{eq:g1(1)}\\
h_1^{(1)}(x)&=& {2x\over [1-x]_+} + {3\over 2}\delta(1-x).
\label{eq:h1(1)}  
\eeq
Inserting these equations into the defining equation in
(\ref{eq:ope1b}), one obtains
\beq
h_L^{WW(1)}(x)&=&3x + 4x{\rm ln} {1-x\over x},
\label{eq:hlww(1)}\\
h_L^{m(1)}(x)&=&
{2\over (1-x)_+}-4x{\rm ln}{1-x\over x}-3 +
{3\over 2}\delta(1-x) + \left( 3x-{3\over 2}\delta(1-x) \right) 
-{1\over 2}\delta(x)\nonumber\\
&=&
{2\over (1-x)_+}-4x{\rm ln}{1-x\over x}-3 +3x
-{1\over 2}\delta(x).
\label{eq:hlm(1)}
\eeq
In the first line of (\ref{eq:hlm(1)}), 
the term $(3x-{3\over 2}\delta(1-x))$ comes from
the self-energy correction, i.e. from expanding
$M=m_q\left[1+\frac{\alpha_S}{2\pi}C_F \frac{3}{2}\ln 
\frac{Q^2}{m_q^2}\right]$ in Eq. (\ref{eq:ope1a}), 
and $-\frac{1}{2}\delta(x)=-\frac{1}{2}g_1(0+)\delta(x)$ 
in $h_L^{m(1)}(x)$ accounts for the second term on 
the righthand side of Eq. (\ref{eq:hlmreg}).  
This term is necessary to reproduce the original
moment relation (\ref{eq:hlm0}).  
We also note that $h_L^{WW(1)}$ does not have any 
singularity at $x=0$ and satisfies
$\int_{0}^1\,dx\,h_L^{WW(1)}(x) =
\int_{0}^1\,dx\,h_1^{(1)}(x)$ as it should.

$h_L^{(1)}(x)$ can be constructed if we know 
the purely twist-3 part $h_L^{3(1)}(x)$ in the one-loop level.   
$h_L^3(x, \mu^2)$
can be written in terms of the quark-gluon light-cone correlation function as
($\xi^2=0$, $\xi^+=0$)\,\cite{BBKT96,BM97,KT99}
\beq
h_L^3(x,Q^2)&=&{iP^+\over M} \int_{-\infty}^\infty {d\xi^- \over 2\pi}
e^{2ixP^+\xi^-} \int_0^1 \,udu\int_{-u}^u\,tdt\nonumber\\
& &\times\la PS_\parallel | \bar{\psi}(-u\xi^-)i\gamma_5\sigma_{\mu\alpha}
gF_\nu^{\ \alpha}(t\xi^-)\xi^\mu\xi^\nu\psi(u\xi^-)|_{Q^2}|PS_\parallel\ra. 
\label{eq:hl3corr}
\eeq
Starting from this expression, one can, 
in principle, obtain $h_L^{3(1)}(x)$. 
Alternatively, and this is the approach that we will
use, one can construct it from already existing
one-loop calculation of its moments.
Taking the $n$-th moment of (\ref{eq:hl3corr})
or from the OPE analysis of $h_L$\,\cite{JJ92}, we have 
\beq
\int_{-1}^1\,dx\,x^nh_L^3(x,Q^2)=\sum_{k=2}^{[(n+1)/2]}
\left( 1 - {2k\over n+2}\right){1\over 2M}
\la PS_\parallel | R_{n,k}(Q^2)|PS_\parallel\ra
\eeq
with
\beq
R_{n,k}(Q^2)={1\over 2} \bar{\psi}(0)\sigma^{\alpha \beta}n_\beta i\gamma_5 
(in\cdot D)^{n-k}igF^\nu_{\ \alpha}n_\nu(in\cdot D)^{k-2}\psi(0)|_{Q^2}
-\left(k \rightarrow n-k+2\right).
\eeq
One-loop renormalization of $h_L$ was completed in \cite{KT95}
and the mixing matrix for the local operators
contributing to the moments of $h_L(x,Q^2)$ was presented.  
In particular, 
matrix elements of $R_{n,k}$ for a massive quark is given in 
eq.(3.18) of \cite{KT95}.
Since $h_L^{(1)}(x)$ for 
the anti-quark is zero, we obtain for the
moment of the quark distribution as
\beq
\int_{0}^1\,dx\,x^nh_L^{3(1)}(x) &=& 2\sum_{l=2}^{[(n+1)/2]}
\left( 1 - {2l\over n+2}\right)\left( {1\over [l-1]_3}
-{1\over [n-l+1]_3}\right)\nonumber\\
&=&{3\over n+1} -{6\over n+2} +{1\over 2} 
\label{eq:hl3(1)mom}
\eeq
for $n\geq 2$ with $[k]_3\equiv k(k+1)(k+2)$.
[The prefactor
in (\ref{eq:hl3(1)mom}) is determined by comparison
with the anomalous dimension for $h_1$ and by noting that the operator
basis for the quark mass operator in \cite{KT95} has a sign opposite
to those for $g_1$.] 
{}From this result and the defining relation for the lowest
two moments of $h_L^3$,
(\ref{eq:hl3mom0}) and 
(\ref{eq:hl3mom1}),
we can construct $h_L^{3(1)}(x)$ as 
\beq
h_L^{3(1)}(x)= 3 -6x + {1\over 2}\delta(1-x) -{1\over 2}\delta(x)
\label{eq:hl3(1)}.  
\eeq
We emphasize that
the $-1/2\delta(x)$ in (\ref{eq:hl3(1)}) is necessary to reproduce the 
$n=0$ moment of $h_L^{3(1)}(x)$. From (\ref{eq:hlww(1)}), 
(\ref{eq:hlm(1)}) and (\ref{eq:hl3(1)}), one obtains 
\beq
h_L(x,Q^2) = \delta(1-x) +
{\alpha_S\over 2\pi}{\rm ln}{Q^2\over m_q^2}C_F
\left[{2\over [1-x]_+} +{1\over 2}\delta(1-x) -\delta(x)\right].
\label{eq:hlx}
\eeq
We remark that
the above calculation indicates
that the $\delta(x)$ term appears not only in
$h_L^{m}$ but also in $h_L^{3}$.  
Furthermore they do not cancel but add up to
give rise to $-\delta(x)$ in $h_L(x,Q^2)$ itself.  

In the next subsection we will confirm Eq. 
(\ref{eq:hlx}) through
a direct calculation of $h_L(x,Q^2)$ for a quark.

\subsection{Light-cone calculation of $h_L(x,Q^2)$}
In order to illustrate the physical origin of the $\delta(x)$ terms in
$h_L(x)$, and in order to develop a more general and convenient procedure
for calculating such terms, we shall now evaluate $h_L(x)$ using
time-ordered light-front (LF) perturbation theory.
The general method has been outlined in Ref. \cite{HZ} and we
will restrict ourselves here to the essential steps only.

There are two equivalent ways to perform time-ordered LF perturbation
theory: one can either work with the LF Hamiltonian for QCD and perform
old-fashioned perturbation theory (the method employed in Ref. \cite{HZ}), 
or one can start from Feynman perturbation theory and integrate over the
LF-energy $k^-$ first. We found the second approach more convenient for
the one-loop calculation of $h_L(x)$ and this is what we will use in the 
following.

In LF gauge, $A^+=0$, parton distributions can be expressed
in terms of LF momentum densities
($k^+$-densities). Therefore, one finds for a parton distribution,
characterized by the Dirac matrix $\Gamma$
at ${\cal O}(\alpha_S)$ and for $0<x<1$
\beq
f_\Gamma(x) \ub(p)\Gamma u(p)= -ig^2 \ub(p)
\int \frac{d^4k}{(2\pi)^4}
\delta\left(x-\frac{k^+}{p^+}\right)
\gamma^\mu \frac{1}{\kslash-m_q+\ieps}
\Gamma \frac{1}{\kslash-m_q+\ieps}\gamma^\nu u(p)
D_{\mu \nu}(p-k),\nonumber\\
\label{eq:lf1}
\eeq
where
\beq
D_{\mu \nu}(q)
=\frac{1}{q^2+\ieps}\left[
g_{\mu \nu}-\frac{q_\mu n_\nu+n_\mu q_\nu}{qn}
\right]
\eeq
is the gauge field propagator in LF gauge, and
$n^\mu$ is a light-like vector such that
$nA=A^+ \sim\left(A^0+A^3\right)/\sqrt{2}$ for any
four vector $A^\mu$.

The $k^-$ integrals in expressions like Eq.
(\ref{eq:lf1}) are performed using Cauchy's theorem,
yielding for $0<k^+<p^+$
\beq
-i\int \frac{dk^-}{2\pi}
\frac{1}{\left(k^2-m_q^2+\ieps\right)^2}
\frac{1}{(p-k)^2+\ieps}
&=& \frac{1}{(2k^+)^2}\frac{1}{2(p^+-k^+)}
\frac{1}{\left(p^--\frac{m_q^2+\kp^2}{
2k^+} - \frac{ (\pp-\kp)^2}{2(p^+-k^+)}
\right)^2}\nonumber\\
& &\stackrel{\kp \rightarrow \infty}
{\longrightarrow}
\frac{1}{2p^+} \frac{1-x}{\kp^4},
\eeq
where we used $k^+=xp^+$.
In order to integrate all terms in Eq. (\ref{eq:lf1}) over $k^-$,
Cauchy's theorem is used to replace any factors of 
$k^-$ in the numerator of Eq. (\ref{eq:lf1}) 
containing $k^-$ by their on-shell value at the pole 
of the gluon propagator
\beq
k^- \longrightarrow \tilde{k}^- \equiv
p^--\frac{(\pp-\kp)^2}{2(p^+-k^+)}
\label{eq:ktilde}
\eeq
In the following we will focus on the UV divergent 
contributions to the parton distribution only. This 
helps to keep the necessary algebra at a reasonable level
because we restrict 
ourselves to the leading behavior in $\kp$, which
arises from  those terms in the numerator of Eq.(\ref{eq:lf1}) that 
are quadratic in $\kp$, and therefore give rise to a
logarithmically divergent $\kp$ integral. The
transverse momentum cutoff can be replaced
by $Q^2$ in these expressions.
The rest of the calculation is just tedious algebra
and we omit the intermediate steps here.

We find for $0<x<1$ to ${\cal O}(\alpha_S)$
\beq
h_L(x,Q^2)= \frac{\alpha_S}{2\pi}C_F 
\ln \frac{Q^2}{m_q^2} \frac{2}{[1-x]_+},
\label{eq:hL1}
\eeq
where the usual $+$-prescription for $\frac{1}{[1-x]_+}$ applies at
$x=1$, i.e.  $\frac{1}{[1-x]_+}=\frac{1}{1-x}$ for $x<1$ and
$\int_0^1dx\frac{1}{[1-x]_+}=0$.
Furthermore, $h_L(x)=0$ for $x<0$, since anti-quarks do not occur in the 
${\cal O}(\alpha_S)$ dressing of a quark. In addition to Eq. (\ref{eq:hL1}), 
there is also an explicit $\delta(x-1)$ contribution at $x=1$. These
are familiar from twist-2 distributions, where they reflect the fact
that the probability to find the quark as a bare quark is less than one
due to the dressing with gluons. For higher-twist
distributions, the wave function renormalization contributes is 
$\frac{\alpha_S}{2\pi}C_F \ln \frac{Q^2}{m_q^2}
\frac{3}{2}\delta(x-1)$. The same wave function renormalization
also contributes at twist-3.
However, for all higher twist distributions there is an additional
source for $\delta(x-1)$ terms
which has, in parton language, more the appearance
of a vertex correction, 
but which
arises in fact from the gauge-piece of self-energies connected to the
vertex by an `instantaneous fermion propagator' $\frac{\gamma^+}{2p^+}$.
For $g_T(x,Q^2)$ these have been calculated in Ref.
\cite{HZ} where they give an additional
contribution $-\frac{\alpha_S}{2\pi}C_F \ln \frac{Q^2}{m_q^2}
\delta(x-1)$, i.e. the total contribution at $x=1$ for $g_T(x,Q^2)$ was
found to be $\frac{\alpha_S}{2\pi}C_F \ln \frac{Q^2}{m_q^2}
\frac{1}{2}\delta(x-1)$. We found the same $\delta(x-1)$ terms also
for $h_L(x,Q^2)$.
\footnote{Physically, the reason why the wave 
function renormalization contribution depends on the 
twist is that in LF gauge, different components of 
the fermion field aquire different wave function 
renormalization. However, since all twist-3 parton
distributions involve one LC-good and one LC-bad 
component, it is natural to find the same wave 
function renormalization for all three twist-3 
distributions.}
Combining the $\delta(x-1)$ piece with Eq. 
(\ref{eq:hL1}) we thus find
\beq
h_L(x,Q^2) = \delta(x-1) +
\frac{\alpha_S}{2\pi}
C_F\ln\frac{Q^2}{m_q^2}\left[\frac{2}{[1-x]_+} + 
\frac{1}{2}\delta(x-1)\right]
\quad \quad \mbox{for}\quad x>0
.\label{eq:hlcan}
\eeq
Comparing this result with the well known result for 
$h_1$\,\cite{AM90}
\beq
h_1(x,Q^2) = \delta(x-1)
+ \frac{\alpha_S}{2\pi} \ln \frac{Q^2}{m_q^2} C_F
\left[\frac{2x}{[1-x]_+} + 
\frac{3}{2}\delta(x-1)\right],
\eeq
one realizes that
\beq
\lim_{\varepsilon \rightarrow 0}
\int_\varepsilon^1 dx \left[
h_L(x,Q^2)-h_1(x,Q^2\right]
=   \frac{\alpha_S}{2\pi} \ln \frac{Q^2}{m_q^2} C_F
\neq 0,
\eeq
i.e. if one excludes the possibly
problematic region $x=0$, then the $h_L$-sum rule
\cite{mb:nagoya} is 
violated already for a quark dressed with gluons 
at order ${\cal O}(\alpha_S)$.

In the above calculation, we carefully avoided
the point $x=0$.
For most values of $k^+$, the denominator in
Eq. (\ref{eq:lf1}) contains three powers of 
$k^-$ when $k^-\rightarrow \infty$.
However, when $k^+=0$, $k^2-m_q^2$ becomes
independent of $k^-$ and the denominator in
Eq. (\ref{eq:lf1}) contains only one power of
$k^-$. Therefore, for those terms in the numerator
which are linear in $k^-$, \footnote{This is the 
highest 
power of $k^-$ that can appear in the numerator
for twist three distributions.}
the $k^-$-integral diverges linearly.
Although this happens only for a point of measure
zero (namely at $k^+=0$), a linear divergence is indicative
of a singularity of $h_L(x,Q^2)$ at that point.
\footnote{There is another point, $k^+=p^+$, where
a similar divergence occurs, but the latter is only logarithmic.}
To investigate the $k^+\approx 0$ singularity in
these terms further, we consider
\beq
f(k^+,\kp)&\equiv& \int dk^-
\frac{k^-}{(k^2-m_q^2+\ieps)^2}
\frac{1}{(p-k)^2+\ieps}
\nonumber\\
&=&
 \int dk^-
\frac{\tilde{k}^-\quad+\quad 
\left(k^--\tilde{k}^-\right)}
{(k^2-m_q^2+\ieps)^2\left[(p-k)^2+\ieps\right]} 
\nonumber\\
&=&f_{can.}( k^+,\kp) + f_{sin.}(k^+,\kp),
\label{eq:fsing1}
\eeq
where the `canonical' piece $f_{can.}$
is obtained by
substituting for $k^-$ its on energy-shell value $\tilde{k}^-$
(\ref{eq:ktilde})
(the value at the pole at $(p-k)^2=0$)
\beq
\tilde{k}^- \equiv p^- -\frac{(\pp-\kp)^2}{2(p^+-k^+)}.
\eeq
For $k^+=xp^+\neq 0$, it is only this canonical
piece which contributes. To see this, we note that
\beq
k^--\tilde{k^-} = - \frac{(p-k)^2}{2(p^+-k^+)},
\label{eq:trick}
\eeq
and therefore
\beq
f_{sin}(k^+,\kp)&=& \int dk^- \frac{k^--\tilde{k}^-}{(k^2-m_q^2+\ieps)^2}
\frac{1}{(p-k)^2+\ieps}
\nonumber\\
&=& \frac{1}{2(p^+-k^+)}\int dk^- \frac{1}{(k^2-m_q^2+\ieps)^2}
\label{eq:sin}.
\eeq
Obviously\cite{CY}
\beq
\int dk^- \frac{1}{(2k^+k^--\kp^2-m_q^2+\ieps)^2}=0
\eeq
for $k^+\neq 0$ because then one can always avoid
enclosing the pole at $k^-=\frac{m_q^2+\kp^2-\ieps}
{2k^+}$ by closing the contour in the appropriate 
half-plane of the complex $k^--plane$.
However, on the other hand
\beq
\int dk^+ dk^- \frac{1}{(2k^+k^--\kp^2-m_q^2+\ieps)^2}
= \int d^2k_L \frac{1}{(k_L^2-\kp^2-m_q^2+\ieps)^2}
=\frac{i\pi}{\kp^2+m_q^2}
\eeq
and therefore
\beq
f_{sin}(k^+,\kp) = {1\over 2p^+} \frac{i\pi\delta(k^+)}{\kp^2+m_q^2}.
\label{eq:delta}
\eeq
Upon collecting all terms $\propto k^-$ in the
numerator of Eq. (\ref{eq:lf1}), and applying Eq. 
(\ref{eq:delta}) to those terms we find after some algebra those 
terms in $h_L(x,Q^2)$ that are singular at $x=0$
\beq
h_{L,sin}(x,Q^2)=
-\frac{\alpha_S}{2\pi} \ln \frac{Q^2}{m_q^2} C_F \delta (x).
\eeq
Together with Eq. (\ref{eq:hlcan}), this gives our 
final result for $h_L$, up to ${\cal O}(\alpha_S)$, 
valid also for $x=0$
\beq
h_L(x,Q^2) = \delta(x-1) +
\frac{\alpha_S}{2\pi}
C_F\ln\frac{Q^2}{m_q^2}\left[-\delta(x)+
\frac{2}{[1-x]_+} + 
\frac{1}{2}\delta(x-1)\right]
.\label{eq:hl}
\eeq
As expected, $h_L$ from Eq. (\ref{eq:hl}) does now satisfy
the $h_L$-sum rule, provided of course the origin is included 
in the integration.

This result is important for several reasons. First of all
it confirms our result for $h_L(x,Q^2)$ as determined from the
moment relations. Secondly, it provides us with a 
method for
calculating these $\delta(x)$ terms and thus
enabling us to address the issue of validity of the naive
sum rules more systematically. And finally, it shows that
there is a close relationship between these $\delta(x)$ terms
and the infamous zero-modes in LF field theory \cite{mb:adv}.

While we were completing the manuscript for this paper, we
learned of Ref. \cite{metz}, where canonical Hamiltonian
light-cone perturbation theory is used to calculate $h_L(x)$.
For $x\neq 0$ the result obtained in Ref. \cite{metz} agrees
with ours which provides an independent check of the formalism 
and the algebra.  However, the canonical light-cone perturbation
theory used in Ref. \cite{metz} is not adequate for studying the
point $x=0$.  From the smooth behaviour of $h_L(x)$ {\it near} 
$x=0$ the authors of Ref. \cite{metz} conclude that the sum
rule for the parton distribution $h_L(x)$ is violated to
${\cal O}(\alpha_S)$.  Our explicit calculation 
for $h_L(x)$ not only proves that the sum rule for $h_L(x)$
is {\it not} violated to this order if the point $x=0$ is properly 
included, but also shows that it is incorrect to draw conclusions 
{}from smooth behaviout near $x=0$ about the behaviour at $x=0$.

\section{$e(x,Q^2)$}
\setcounter{equation}{0}
The other chiral-odd twist-3 distribution $e(x,Q^2)$
is also expected to satisfy a simple operator sum rule
\beq
\int_{-1}^1dx e(x,Q^2) = \frac{1}{2M}\langle P|\left.
\bar{\psi}(0)\psi(0)\right|_{Q^2}|P\rangle,
\label{eq:sigma}
\eeq
which follows trivially by integrating (\ref{eq3}) over $x$.
Because of our results from above for $h_L(x,Q^2)$, we are now of course
more cautious and address in the following the issue whether
Eq. (\ref{eq:sigma}) is also valid if the origin is excluded from
the region of integration.
Again, we will consider a massive quark to ${\cal O}(\alpha_S)$

\subsection{Constructing $e(x,Q^2)$ from its moments}
The OPE analysis of (\ref{eq3}) decomposes
$e(x)$ into the twist-2 contribution and the purely
twist-3 piece $e^3(x)$ as
\beq
{\cal M}_n[e]={\cal M}_n[e^3] + {m_q\over M}{\cal M}_{n-1}[f_1],\quad
(n\geq 1)
\label{eq:emom}
\eeq
with the relation (\ref{eq:sigma}).  This moment 
relation defines the decomposition
\beq
xe(x,Q^2)=xe^3(x,Q^2) + {m_q\over M}f_1(x,Q^2).
\label{eq:edecomp}
\eeq
where we multiplied $x$ to regularize any possible 
$\delta(x)$ contributions, which will be specified 
later. The one loop calculation of $e(x,Q^2)$ for a 
massive quark yields
\beq
e(x,Q^2) = \delta(1-x) + {\alpha_S\over 2\pi}C_F{\rm ln}{Q^2\over m_q^2}
e^{(1)}(x).
\eeq
The lowest moment of $e^{(1)}(x)$ can be obtained 
directly from the $\sigma$-term relation
(\ref{eq:sigma})
\beq
{\cal M}_0[e]={1\over 2M}\la P|\bar{\psi}\psi|P\ra = {1\over 2M}{\partial 
\over \partial m_q}M^2 = {\partial M\over \partial m_q} 
=1+ {\alpha_S\over 2\pi}C_F\times {3\over 2}\ln 
\frac{Q^2}{m_q^2},
\eeq
i.e. the lowest moment of $e^{(1)}$ reads
\beq
{\cal M}_0[e^{(1)}]= {3\over 2}.
\label{eq:e(1)0}
\eeq
Corresponding to (\ref{eq:emom}), we have
\beq
{\cal M}_n[e^{(1)}]={\cal M}_n[e^{3(1)}] + 
{1\over M}{\cal M}_{n-1}\left[\left(m_qf_1\right)^{(1)}\right],\quad
(n\geq 1)
\label{eq:e(1)mom}
\eeq
Our problem is to construct $e^{(1)}(x)$ for a massive quark
{}from (\ref{eq:e(1)0}) and
(\ref{eq:e(1)mom}).
In principle we can obtain
the purely twist-3 piece $e^{3(1)}$ from the one-loop calculation
starting from the correlation function\,
\cite{BBKT96,BM97,KT99}:
\beq
e^3(x,Q^2)={P^+\over M}\int\,{d\xi^-\over 2\pi}
e^{2ixP\cdot\xi}\int_0^1\,du \int_{-u}^u\,
dt \la P| \bar{\psi}(-u\xi)\sigma^{\mu\alpha}gF_{\nu\alpha}(t\xi)
\xi_\mu\xi^\nu\psi(u\xi)|_{Q^2}| P \ra.
\eeq
However, we again make a short cut to get it from the moment.  
The moment of $e^{3(1)}(x)$ for a massive quark
is given in \cite{KN97} as a part of the mixing matrix 
in the context of the renormalization.
{}From eqs.(4.2) and (3.18) of \cite{KN97} we have for the
$n$-th moment of $e^{3(1)}$ ($n\geq 1$):
\beq
\int_{-1}^1\,dx\,x^ne^{3(1)}(x)&=&\left\{
\begin{array}{ll}
\displaystyle{
2\sum_{l=2}^{n/2}\left[ {1\over [l-1]_3} + {1\over [n-l+1]_3} \right]
+{2\over [n/2]_3}
},&\qquad (n: even)\\[15pt]
\displaystyle{
2\sum_{l=2}^{(n+1)/2}\left[ {1\over [l-1]_3} + {1\over [n-l+1]_3} \right]},
&\qquad (n: odd)
\end{array}\right.\nonumber\\
&=& {1\over 2} -{1\over n} +{1\over n+1}
\eeq
which gives 
\beq
xe^{3(1)}(x) = x{1\over 2 }\delta(1-x) -1 +x.
\eeq
Together with the twist-2 contribution
\beq
{1\over M}\left(m_qf_1\right)^{(1)}(x)= {1+x^2\over [1-x]_+},
\eeq
one obtains
\beq
e^{(1)}(x) =  {2\over [1-x]_+} + {1\over 2}
\delta(1-x) 
\quad\quad \quad \mbox{for} \quad x>0.
\label{eq:e(1)final}
\eeq
Note that $1/x$ singularity in $e^{3(1)}(x)$ and $f_1^{(1)}/x$ cancel each 
other in $e^{(1)}$, and $e^{(1)}(x)$ itself is 
integrable at $x=0$. 
In order to satisfy the $\sigma$-term sum rule (\ref{eq:e(1)0})
one needs to introduce $\delta(x)$ term.
Accordingly the final result, which
satisfies the moment relations both for $n=0$ and
$n\neq 0$, reads  
\beq
e^{(1)}(x) =  {2\over [1-x]_+} + {1\over 2}
\delta(1-x) + \delta(x).
\label{eq:e(1)final2}
\eeq
This result clearly indicates that the sum rule (\ref{eq:sigma})
is satisfied only by including the point $x=0$ 
in the integration.

\subsection{Light-cone calculation}
We start from Eq. (\ref{eq:lf1}) with $\Gamma=1$
\beq
e(x) \ub(p)\Gamma u(p)= -ig^2 \ub(p)
\int \frac{d^4k}{(2\pi)^4}
\delta\left(x-\frac{k^+}{p^+}\right)
\gamma^\mu \frac{1}{\kslash-m_q+\ieps}
\frac{1}{\kslash-m_q+\ieps}\gamma^\nu u(p)
D_{\mu \nu}(p-k).\nonumber\\
\label{eq:lfe}
\eeq
the numerator algebra in Eq. (\ref{eq:lfe}) yields
\beq
& &\!\!\!
\bar{u}(p)\gamma_\mu\left(\kslash+m_q\right)^2 
\gamma_nu u(p)\left[g^{\mu \nu}-\frac{(p-k)^\nu n^\mu
+ (p-k)^\mu n^\nu}{p^+-k^+}\right]
\nonumber\\
&=& 2\bar{u}(p)\left[ k^2+m_q^2+2\frac{k^2p^+-m_q^2k^+}
{p^+-k^+} -2m_q\kslash\right]u(p)
\label{eq:edirac}.
\eeq
In order to determine the `canonical' part of $e(x,Q^2)$
we use again contour integration, picking up the
pole at $k^-=\tilde{k}^-$ for $0<k^+<p^+$, which
allows us to replace
$
k^2\rightarrow -\frac{{\bf k}_\perp^2}{1-x}
$
and 
$p^+k^-\rightarrow -\frac{{\bf k}_\perp^2}{2(1-x)}$,
where we kept only the leading terms in 
${\bf k}_\perp$. After some algebra this yields
for $0<x<1$
\beq
e^{(1)}(x)=\frac{2}{1-x}
\eeq 
in agreement with the result obtained from the
moments. For the wave function renormalization
(the coefficient in front of $\delta(x)$ the
same coefficient is obtained as for $h_L(x,Q^2)$ and
the details will be omitted here.

Finally, we focus on possibly singular terms near
$x=0$. The only numerator term involving $k^-$ in
Eq. (\ref{eq:edirac}), which is not multiplied by
$k^+$ appears in $-4m_q\bar{u}(p)\kslash u(p)$.
Upon repeating the same steps as in Section 3, i.e.
isolating the singular piece by adding and
subtracting $\tilde{k}^-$,  canceling
$k^--\tilde{k}^-$ against the gluon propagator, and making use of Eq. 
(\ref{eq:delta}) one finds 
\beq
e^{(1)}(x)_{sin}= \delta(x).
\eeq
Collecting all terms we thus find
\beq
e^{(1)}(x) =  {2\over [1-x]_+} + {1\over 2}
\delta(1-x) + \delta(x).
\label{eq:e(1)LF},
\eeq
which completely agrees with the result from above
(\ref{eq:e(1)final2})

\section{$g_T(x,Q^2)$}
\subsection{Possibility of $\delta(x)$ in $g_T$} 
\setcounter{equation}{0}
{}From a practical point of view, $g_T(x,Q^2)$ is the most important among
the twist-3 distributions, because it is the least
difficult to measure experimentally.
For this reason, $g_T(x,Q^2)$ has been the subject of many studies in 
the literature. 

We again start with the decomposition of $g_T(x)$ into the 
Wandzura-Wilcek term $g_T^{WW}(x)$ \cite{WW77}, the term proportional to
the quark mass $g_T^m(x)$, and the purely twist-3 part $g_T^3(x)$:
\footnote{Here and below in this subsection 
we omit the scale dependence of the
distribution functions for simplicity.}  
\beq
g_T(x)&=& g_T^{WW}(x)+g_T^m(x)+g_T^3(x)\label{eq:gtdecom}\\
&=&\left\{
\begin{array}{cc}
\displaystyle{
\int_x^1dy \frac{g_1(y)}{y} + \frac{m_q}{M}\left[
\frac{h_1(x)}{x}-\int_x^1dy\frac{h_1(y)}{y^2}\right]
+g_T^3(x)} & \qquad(x>0),\\[20pt]
\displaystyle{
-\int^x_{-1}\,dy \frac{g_1(y)}{y} + \frac{m_q}{M}\left[
\frac{h_1(x)}{x}+\int^x_{-1}\,dy\frac{h_1(y)}{y^2}\right]
+g_T^3(x)} & \qquad(x<0).
\end{array}
\right.
\label{eq:gt}
\eeq
This decomposition is obtained by the analysis
of the correlation function using QCD equation of motions.  
The corresponding moment relations are
\beq
{\cal M}_n[g_T] &=& {1\over n+1}{\cal M}_{n-1}[g_1] +
{n\over n+1}{m_q\over M}{\cal M}_{n-1}[h_1] + {\cal M}_n[g_T^3],
\qquad ( n\geq 1)
\label{eq:gtn}\\
{\cal M}_0[g_T] &=& {\cal M}_0[g_1],
\label{eq:BCSR}
\eeq
with 
\beq
{\cal M}_0[g_T^3] =0,
\label{eq:gt30}
\eeq
where we assumed that
$g_T^3(x)$ itself is integrable at $x=0$.
The relation (\ref{eq:BCSR}) is known as the Burkhardt-Cottingham 
(BC) sum rule\,\cite{BC}. 
We note (\ref{eq:BCSR}) and (\ref{eq:gt30})
implies 
\beq
{\cal M}_0[g_T^m]=0.
\label{eq:gtm0}
\eeq
In the above discussion we again assumed separate relations 
(\ref{eq:gt30}) and (\ref{eq:gtm0}) instead of 
${\cal M}_0[g_T^m+g_T^3]=0$.  This is justified as long as
$h_1(0\pm)$ is finite. (See below.)

Following the same argument leading to (\ref{eq:hlmreg}) for $h_L^m(x)$,
we have 
\beq
\int_{-1}^{0-}dx g_T^m(x)+ \int_{0+}^1dx g_T^m(x) =
{m_q\over M} \left(h_1(0+)-h_1(0-)\right).
\label{eq:gtmsing}
\eeq
This relation together with (\ref{eq:gtm0}) implies 
$g_T^m(x)$ has a singularity at $x=0$ as
\beq
\left. 
g_T^m(x) =
g_T^m(x)\right|_{reg} - \frac{m_q}{M}\left(h_1(0+)-h_1(0-)
\right)\delta(x),
\label{eq:gtmreg}
\eeq
where $g_T^m(x)|_{reg}$ is a part obtained from the integral
(\ref{eq:gt}) at $x>0$ and $x<0$ and is regular at $x=0$.  
This relation shows that if
$h_1(0-)\equiv  -\bar{h}_1(0+) \neq h_1(0+)$ then
the Burkhardt-Cottingham relation would be violated 
if data is
taken only for nonzero $x$ --- unless of course 
there is another $\delta(x)$ contribution to 
$g_T^3(x)$ which happens to cancel exactly the one 
in $g^m_T$.  
Perturbation theory predicts at small $x$ that 
$h_1(x) \sim x$ 
in the leading order as is seen from (\ref{eq:h1(1)})
and $h_1(x)-h_1(-x)\sim {\rm const.}$ at the next-to-leading 
order\,\cite{HKK}.  BFKL approach gives $h_1(x)\sim {\rm const.}$
at small $x$\,\cite{KMSS}.  Therefore there is a possibility that 
the $\delta(x)$ term in (\ref{eq:gtmreg}) survives
and hence a seeming violation of BC sum rule.  

In the next subsections, we will calculate $g_T(x,Q^2)$ for a massive quark
to ${\cal O}(\alpha_S)$ in order to look for (possibly) another origin of
the $\delta(x)$ contribution.  

\subsection{Constructing $g_T(x,Q^2)$ from its moments to
${\cal O}(\alpha_S)$}

One loop calculation of $g_T(x)$ for a massive quark takes the form of
\beq
g_T(x,Q^2) = \delta(1-x) + {\alpha_S\over 2\pi}C_F{\rm ln}{Q^2\over m_q^2}
g_T^{(1)}(x).
\eeq
Corresponding to the decomposition (\ref{eq:gtdecom}) we write
\beq
g_T^{(1)}(x) = g_T^{WW(1)}(x)+g_T^{m(1)}(x) + g_T^{3(1)}(x).
\eeq
Inserting the expression (\ref{eq:g1(1)}) for $g_1^{(1)}$  and
(\ref{eq:h1(1)}) for 
$h_1^{(1)}$ into (\ref{eq:gt}), and also taking into account
the self-energy correction, we have
\beq
g_T^{WW(1)}(x)+g_T^{m(1)}(x) = {\rm ln}x + {1\over 2} + x +{2\over [1-x]_+}.
\label{eq:gtwwm(1)}
\eeq
This result has no $\delta(x)$ term, which is simply because
$h_1(0^+)=h_1(0^-)=0$ in the one-loop calculation.
(See (\ref{eq:gtmreg}).)

As in the case of $h_L^3(x)$ and $e^3(x)$, we construct
$g_T^{3(1)}$ from its moment for a massive quark. 
{}From eqs.(5) and (18) of \cite{KYTU97}, we have for the $n$-th moment of
$g_T^{3(1)}(x)$ as
\beq
\int_{-1}^1\,dx\,x^ng_T^{3(1)}(x)&=&
\sum_{l=1}^{n-2}(n-l){2\over (n+1)l(l+1)(l+2)}
\nonumber\\
&=&{1\over 2} -{3\over 2(n+1)} +{1\over (n+1)^2}. 
\eeq
This moment relation is originally given for $n\geq 1$, but it
also gives zero for $n=0$, which is consistent with (\ref{eq:gt30}).  
Accordingly no $\delta(x)$ term is necessary to obtain 
\beq
g_T^{3(1)}(x)= {1\over 2}\delta(1-x) -{3\over 2} -{\rm ln}\,x.
\label{eq:gt3(1)}
\eeq
From
(\ref{eq:gtwwm(1)}) and (\ref{eq:gt3(1)}), we get
\beq
g_T(x,Q^2)=\delta(1-x)+{\alpha_S\over 2\pi}C_F
\ln \frac{Q^2}{m_q^2}
\left[ {1+2x -x^2\over [1-x]_+} + {1\over 2}\delta(1-x)\right].
\eeq

\subsection{LC calculation}

We calculated $g_T(x,Q^2)$, using using the same LF pQCD techniques 
that we used to study $h_L(x,Q^2)$ and $e(x,Q^2)$. The algebraic steps
involved are rather lengthy and we therefore present only the
final result here, which reads
\beq
g_T(x,Q^2) = \delta(1-x) +
\frac{\alpha_S}{2\pi}
C_F\ln\frac{Q^2}{m_q^2}\left[2x+\frac{1+x^2}{[1-x]_+} + 
\frac{1}{2}\delta(1-x)\right]
\eeq
without any $\delta(x)$ term. This result completely agrees with
the findings from Ref. \cite{HZ}, which confirms our formalism.

Even though the numerator for $\Gamma=\gamma_\perp \gamma_5$
in Eq. (\ref{eq:lf1}) contains $k^-$, those terms turn out to
be multiplied by at least one power of $k^+$. Since $x\delta(x)=0$,
there are no $\delta(x)$ terms in $g_T(x,Q^2)$ to ${\cal O}(\alpha_S)$.
However, we do not have a simple explanation (other than
working out the numerator algebra) as to why factors
of $k^-$ in the numerator algebra are always accompanied by at least
one power of $k^+$ for $g_T(x,Q^2)$ and not for $h_L(x,Q^2)$ and $e(x,Q^2)$ in
this one loop calculation.

\section{Summary}
\setcounter{equation}{0}
We have investigated sum-rules for twist-3 distributions in QCD,
and found examples where these sum-rules are violated if the point 
$x=0$ is not properly included.

For a massive quark, to ${\cal O}(\alpha_S)$ we found
\beq
g_T(x,Q^2) &=& \delta(x-1) +
\frac{\alpha_S}{2\pi}
C_F\ln\frac{Q^2}{m_q^2}\left[2x+\frac{1+x^2}{[1-x]_+} + 
\frac{1}{2}\delta(x-1)\right]
\nonumber\\
h_L(x,Q^2) &=& \delta(x-1) +
\frac{\alpha_S}{2\pi}
C_F\ln\frac{Q^2}{m_q^2}\left[-\delta(x)+\frac{2}{[1-x]_+} + 
\frac{1}{2}\delta(x-1)\right]
\nonumber\\
e(x,Q^2) &=& \delta(x-1) + \frac{\alpha_S}{2\pi}
C_F\ln\frac{Q^2}{m_q^2}\left[\delta(x)+\frac{2}{[1-x]_+} + 
\frac{1}{2}\delta(x-1)\right]
.\label{eq:higherT}
\eeq
At ${\cal O}(\alpha_S)$ neither $h_L(x,Q^2)$ nor $e(x,Q^2)$ satisfy
their respective sum rule if one excludes the origin from
the region of integration (which normally happens in 
experimental attempts to verify a sum rule). $g_T(x,Q^2)$ is the
only exception and its sum-rule is satisfied even when the
origin is not included.

Of course, QCD is a strongly interacting theory and parton
distribution functions in QCD are nonperturbative observables.
Nevertheless, if one can show that a sum rule fails already
in perturbation theory, then this is usually a very strong
indication that the sum rule also fails nonperturbatively
(while the converse is often not the case!).

{}From the QCD equations of motion, we were able to show 
nonperturbatively that
\footnote{The only assumption that we made is that twist-2
distributions do not contain a $\delta$-function at $x=0$.}
the difference between $h_L(x,Q^2)$ and $h_L^3(x,Q^2)$ contains
a $\delta$ function at $x=0$
\beq
\left[h_L(x,Q^2)-h_L^3(x,Q^2)\right]_{singular}=
-\frac{m_q}{2M}\left(g_1(0+,Q^2)-g_1(0-,Q^2)\right)\delta(x).
\eeq
Since $g_1(0+,Q^2)-g_1(0-,Q^2)\equiv \lim_{x\rightarrow 0}
g_1(x,Q^2)-\bar{g}_1(x.Q^2)$ seems to be nonzero (it may even 
diverge\footnote{
In the next-to-leading order QCD for a quark, 
$\lim_{x\to 0}g_1(x)-\bar{g}_1(x)$ is 
logarithmically divergent\,\cite{g1nlo}.}), 
one can thus conclude that either $h_L(x,Q^2)$ or
$h_L^3(x,Q^2)$ or both do contain such a singular term.

We checked the validity of this relation to ${\cal O}(\alpha_S)$
and found that, to this order, both $h_L^3$ and $h_L$
contain a term $\propto \delta(x)$. We also verified that
even though the sum rule for $h_L(x)$ and $e(x)$ are violated if
$x=0$ is not included, the sum rules for all three twist 3
parton distributions are still satisfied to ${\cal O}(\alpha_S)$
if the contribution from $x=0$ (the $\delta(x)$ term) is included.

\subsection*{Acknowledgments}
M.B. was supported by a grant from the DOE (FG03-95ER40965), 
through Jefferson Lab by contract DE-AC05-84ER40150 under 
which the Southeastern Universities Research Association 
(SURA) operates the Thomas Jefferson National Accelerator 
Facility.  Y.K. is supported by the Grant-in-Aid for Scientific
Research (No. 12640260) of the Ministry of Education, Culture, Sports,
Science and Technology (Japan).    
We are also greatful to JSPS for the Invitation Fellowship
for Research in Japan (S-00209) which made it possible to 
materialize this work.

\end{document}